\documentclass[sigconf, preprint]{acmart}

\copyrightyear{2023}
\acmYear{2023}
\setcopyright{acmlicensed}\acmConference[\textbf{Preprint} to appear in IEEE/ACM Workshop on the Internet of Safe Things 2023.\\
CPS-IoT Week Workshops'23]{}{May 9--12,
2023}{San Antonio, TX, USA}
\acmBooktitle{
Cyber-Physical Systems and Internet of Things Week 2023\textbf{-- this is a preprint version}, May 9--12, 2023, San Antonio, TX, USA}
\acmPrice{15.00}
\acmDOI{10.1145/3576914.3587485}
\acmISBN{979-8-4007-0049-1/23/05}

\setcopyright{none}

\usepackage{graphicx}
\usepackage{tabularx}
\usepackage{float}
\usepackage{hyperref}
\usepackage{longtable}
\usepackage{multirow}
\usepackage{array}
\usepackage[justification=centering]{caption}
\usepackage{tikz}
\usetikzlibrary{positioning}
\usepackage{soul}
\usepackage{hyperref}

\newcommand{\eg}{e.g.,~}
\newcommand{\ie}{i.e.,~}

\newcommand\vul[1]{\textbf{\emph{#1}}}

\usepackage{tablefootnote}

\newcommand{\blind}[1]{BLINDED}

\AtBeginEnvironment{tabular}{\footnotesize}

\begin{document}

\title{
Insecure by Design in the Backbone of Critical Infrastructure
}

\author{Jos Wetzels}
\affiliation{
  \institution{Forescout}
  \country{United States}
}
\email{jos.wetzels@forescout.com}

\author{Daniel dos Santos}
\affiliation{
  \institution{Forescout}
  \country{United States}
}
\email{daniel.dossantos@forescout.com}

\author{Mohammad Ghafari}
\orcid{0000-0002-1986-9668}
\affiliation{
  \institution{TU Clausthal}
  \country{Germany}
}
\email{mohammad.ghafari@tu-clausthal.de}



%
%
\begin{abstract}

We inspected 45 actively deployed Operational Technology (OT) product families from ten major vendors and found that every system suffers from at least one trivial vulnerability. 
We reported a total of 53 weaknesses, stemming from insecure by design practices or basic security design failures.
They enable attackers to take a device offline, manipulate its operational parameters, and execute arbitrary code without any constraint. 
We discuss why vulnerable products are often security certified and appear to be more secure than they actually are, and we explain complicating factors of OT risk management.

\end{abstract}



\maketitle

\pagestyle{plain}

%
%
\section{Introduction}
\label{sec:introduction}

Gartner defines Operational Technology (OT) as ``hardware and software that detects or causes a change, through the direct monitoring and/or control of industrial equipment, assets, processes and events''~\cite{gartner-ot}. 
Industrial Control Systems (ICS) are a subset of OT which often drive mission-critical processes and as such security issues affecting them have profound implications
ICS include Supervisory Control and Data Acquisition (SCADA) systems which remotely control and monitor physical processes via connections to specialized embedded devices such as Remote Terminal Units (RTUs) and Programmable Logic Controllers (PLCs).
These PLCs and RTUs implement control loops and are either directly connected to sensors and actuators or to intermediary field devices implementing dedicated control functionality.
Human Machine Interfaces (HMIs) enable both remote SCADA operators as well as local operators to graphically view and control processes through interaction with PLCs and RTUs.
In addition to SCADA systems, Distributed Control Systems (DCS) offer more turnkey vendor-proprietary ecosystems consisting of HMIs, server applications, and controllers typically tailored to specific industrial segments.

%

The importance of OT security in critical infrastructure cannot be overstated.
Over the past decade, there has been significant efforts in security awareness, system hardening, and product certifications in this domain.
Nevertheless, \emph{it is unknown whether the security posture of OT products in the existing install base has actually improved}.
To clear this, 
we inspected 45 OT product families from ten different major vendors, which are actively deployed across various sectors.
We reverse engineered each product and looked for insecure-by-design practices such as unauthenticated protocols and basic security design failures such as weak cryptography. 
%

We found that every product suffers from at least one trivial vulnerability. 
We reported a total of 53 weaknesses, including several critical issues, with impacts ranging from denial-of-service and configuration manipulation to remote code execution.
%
%
We performed additional Open Source Intelligence (OSINT) and found that a sizable number of these systems are being exposed online.
Worryingly, many of these products are certified but suffer from vulnerabilities that should have been caught in the certification process.
We share our observations and discuss obstacles to improving OT security.



%
%

\section{Methodology}
\label{sec:methodology}


%

We inspected 45 OT product families from ten major vendors, listed in Table~\ref{listofproducts}.
We selected these products based on their prevalence in Forescout customer production environments, which are representative of several big Fortune 1000 corporations and public infrastructure entities.
%
%
We studied a variety of documents, such as vendor manuals and regulatory reports, to understand the technical composition of software and firmware components of each product. 
We compiled a list of explicit and implicit security controls and features for each component. Explicit features include information such as authentication methods, and implicit features include measures that serve security purposes but may not be advertised as such, such as encrypted firmware for vendor IP protection and operational features (\eg physical run/program mode switches) with security consequences.


%
%
We interacted with each device and captured the traffic generated by that interaction to recover the structure and semantics of a protocol to the greatest extent possible.
The structure concerns, \eg division into header/payload and individual field boundaries, whereas the semantics concerns the meaning of particular fields based on a combination of their structural position and/or contextual usage.
We had to understand whether a protocol is conventional or proprietary. 
Conventional protocols are those which rely on open or otherwise well-understood protocols such as SSH, Telnet, FTP, and RPC.
These protocols have well-defined and published structure and semantics, and therefore, their security features are known and documented.
In contrast, understanding any proprietary extensions or modifications to conventional protocols (\ie proprietary protocols) require more investigations than only packet inspection.
%
%
%
Precisely,
packet inspection is limited in a few ways. 
We cannot reconstruct what is not in the capture such as undocumented/rare behavior.
We may need to guess/infer semantics of some fields, which increases the risk of false positives.
Packet inspection can be opaque for traffic that is encrypted, compressed, and traffic that adopts proprietary encoding.
Therefore, we reverse engineered the binaries to relate packet structure and values to semantics. 
We were mainly interested to understand Parsers and Crafters. The former (\ie Parsers) include routines which parse, dispatch, and handle incoming network traffic, whereas the latter (\ie Crafters) comprise routines which compose and format outgoing network traffic.

%
%
Once we obtained a deep understanding of a protocol, we checked the protocol for common weaknesses based on at least the following four aspects:
(i) Authentication and Authorization:
is there an identity and access management process to restrict protocol interactions to proper users and roles?
(ii) Confidentiality: is sensitive data protected from interception?
(iii) Integrity:
is protocol data protected from manipulation?
(iv) Availability:
is it possible to violate system availability through the protocol?
In addition, we enumerated so-called Potentially Dangerous Operations supported by each protocol such as starting/stopping PLCs, firmware upgrades, and logic downloads, and we determined whether the above properties sufficiently protect these operations within the protocol context as well as within the system context (\eg not only firmware update functionality is authenticated in the protocol but also firmware signing is enforced by the device).
We always followed the ``four eyes'' principle, where another person validated every finding.
We eliminated any incorrect conclusion through experimental validation (\eg development of a PoC), and vendor validation where a vendor approves or rejects a finding based on their knowledge of the system.

We followed responsible disclosure practices and reported every issues to vendors.
They disagreed with us in several cases.
We dealt with this by asking for their reasons to dispute the finding.
We then re-evaluated our own report in light of the vendors disagreements, and if we found ourselves aligned with the vendor, we dropped or adjusted the report.
If the report was adjusted, we re-sent it for a new approval round.
Otherwise, if we stood by our original report, we provided our rational and notified the vendor that we see no reason to revise our original report and intend to disclose it as-is.
Typically, this was followed by a conference call where both sides could discuss more the nature of disagreements.

Disagreement occurred in several cases.
We typically reached a mutually acceptable compromise when disagreements were over disclosure timelines or CVSS scores.
We came around to the vendor's point of view in less than five cases.
However,
we were unable to reach an agreement with the vendor for more than ten but less than 15 issues and stood by our original point and went ahead with the disclosure. 
The majority of these disagreements were over issues pertaining to only two vendors. 
In addition, one other vendor did not respond to our disclosures at all and every disclosure for them went exclusively through CISA.

When a vendor refused to acknowledge something is an issue at all, this led to the unfortunate situation of a public CVE without corresponding vendor guidance or acknowledgement.

\begin{table}[]
\caption{List of vendors and product families}
\label{listofproducts}
\centering
\begin{tabular}{|p{0.23\linewidth}|p{0.67\linewidth}|}
\hline
\textbf{Vendor} & \textbf{Product Families} \\ \hline
Bently Nevada   & 3500/3701                           \\ \hline
Emerson         & DeltaV, Ovation, OpenBSI, ControlWave, BB 33xx, ROC, Fanuc/PACsystems                          \\ \hline
Honeywell       & Trend IQ*, Safety Manager, Experion LX, ControlEdge, Saia Burgess PCD                           \\ \hline
JTEKT           & Toyopuc                          \\ \hline
Motorola        & MOSCAD/ACE IP gateway, MDLC, ACE1000                          \\ \hline
Omron           & SYSMAC Cx series, SYSMAC Nx series                          \\ \hline
Phoenix Contact & ProConOS                          \\ \hline
Siemens         & WinCC OA                          \\ \hline
Yokogawa        & STARDOM                          \\ \hline
Schneider Electric        & Modicon Unity                    \\ \hline
\end{tabular}
\end{table}

%
%

\section{Result}
\label{sec:vulnerabilities}


We present the technical composition of software and firmware components to enlighten the capabilities that one should develop to conduct OT vulnerability research.
We report the vulnerabilities found, and estimate where the vulnerable systems are in operation.



\subsection{Technical Information}


%



The compiler signatures and recovered vtable structures showed that the vast majority of software components (\ie 84\%) were written in C++ which is typically more tedious and involved than C or .NET. 
Many software packages heavily used MFC, ATL, COM, RPC, and Qt.
In some cases, a single subsystem implementation (such as a protocol parser and dispatcher) was spread out across multiple DLLs loaded by different processes interacting through IPC.
The other programming languages were C\# (10\%), Delphi (5\%), and lastly Visual Basic (1\%).

Every firmware was exclusively written in a mix of assembly and C(++).
There were no encryption or obfuscation, but they largely relied on proprietary file formats.

We enumerated the CPU architectures and found that ARM (31\%), X86 (26\%), and PowerPC (24\%) are the top most used CPU architectures.
SuperH was used 12\%, and the remaining 7\% belonged to other architectures.

In terms of the underlying operating systems of relevant firmware, the top three were
VxWorks (22\%),
QNX (14\%), and
Linux (13\%).
WinCE had a share of 9\%, followed by 
OS-9 (4\%) and ITRON/T-KERNEL (4\%).
We identified that 11\% rely on a custom OS, and 23\% used other operating systems.
Interestingly, we noticed some significant outliers based either on region (\eg SuperH with OS-9 or ITRON in Asia) or market share (\eg many smaller vendors opted for free or obscure/legacy RTOSes).
%
%
The diversity of OSes shows that reverse engineers face heterogeneous environments, making their efforts more expensive and less scalable.


\subsection{Vulnerabilities}

We found at least one trivial vulnerability in each product and reported a total of 53 distinct CVEs.\footnote{
Two CVEs belong to Schneider Electric products and are not yet fully disclosed, as of the date of finalization of this document on March 10, 2023.
}
More than a third, \ie 21 CVEs, relate to compromise of credentials.
Manipulation is the second most common issue with 18 CVEs, of which the majority (\ie 13 CVEs) are about firmware manipulation.
Remote code execution is the third one with 10 CVEs.
%
%
%
%
%
%
%
%
We recommend to check CVE reports and advisories for an explanation of each vulnerability and remediation.





\vul{Manipulation} allows an attacker to change important aspects of a device or system, such as operational parameters, files stored within the device, the firmware running on the device, or specific configurations of the device. 
%
We identified 16 vulnerabilities with manipulation impact that are present in Table~\ref{manipulation}.
The common root causes were 
a lack of authentication in the respective protocols,
hard-coded credentials, unsigned firmware images, and insecure checksums for integrity checks.

\begin{table}
\centering
\caption{Manipulation}
\label{manipulation}
\begin{tabular}{|p{0.2\linewidth}|p{0.7\linewidth}|}
\hline
\textbf{Vendor} & \textbf{CVEs} \\ \hline \hline
Bently Nevada & CVE-2022-29952 \\ \hline
JTEKT & CVE-2022-29951   \\ \hline  
Yokogawa & FSCT-2022-0039   \\ \hline  
Omron & CVE-2022-31207  \\ \hline
Motorola & CVE-2022-30272, CVE-2022-30276 \\ \hline  
Honeywell & CVE-2022-30313, CVE-2022-30314, CVE-2022-30316, CVE-2022-30317   \\ \hline    
Emerson & CVE-2022-29966, CVE-2022-29957, CVE-2022-30260, CVE-2022-30267, CVE-2022-30262, CVE-2022-30268   \\ \hline  
\end{tabular}
\end{table}

\vul{Compromise of Credentials} allows an attacker to obtain credentials to device functions usually because they are stored or transmitted insecurely.
We identified 20 vulnerabilities with Compromise of Credentials impact, which are listed in Table~\ref{credentials}.
The common root causes were 
transmission of secrets (\eg passwords and tokens) in plain text,
hard-coded credentials even for root users,
weak encryption (\eg deterministic and insecure algorithms),
and default undocumented credentials.

\begin{table}[h]
\caption{Compromise of Credentials}
\label{credentials}
\centering
\begin{tabular}{|p{0.2\linewidth}|p{0.7\linewidth}|}
\hline
\textbf{Vendor} & \textbf{CVEs}                                                                                                                                                  \\ \hline \hline
Emerson         & CVE-2022-29954, CVE-2022-29955, CVE-2022-29959, CVE-2022-29960, CVE-2022-20361, CVE-2022-20362, CVE-2022-20363, CVE-2022-20364, CVE-2022-29965, CVE-2022-30266 \\ \hline
Honeywell       & CVE-2022-30312, CVE-2022-30318, CVE-2022-30320                                                                                                                 \\ \hline
Motorola        & CVE-2022-30270, CVE-2022-30271, CVE-2022-30274, CVE-2022-30275                                                                                                 \\ \hline
Omron           & CVE-2022-31205                                                                                                                                                 \\ \hline
Yokagawa        & CVE-2022-29519, CVE-2022-30997                                                                                                                                 \\ \hline
\end{tabular}
\end{table}


\vul{Authentication Bypass} allows an attacker to bypass existing authentication functions and invoke desired functionality on the target device. 
We identified four vulnerabilities with such impact, which are listed in Table~\ref{authentication}.
The common root causes were 
authentication based on MAC/IP whitelisting,
legacy encryption mode which offers no message integrity, and 
client-side only authentication.

\begin{table}[h]
\caption{Authentication Bypass}
\label{authentication}
\centering
\begin{tabular}{|p{0.22\linewidth}|p{0.68\linewidth}|}
\hline
\textbf{Vendor} & \textbf{CVEs}  \\ \hline \hline
Emerson         & CVE-2022-29961 \\ \hline
Honeywell       & CVE-2022-30319 \\ \hline
Motorola        & CVE-2022-30273 \\ \hline
Siemens         & CVE-2022-33139 \\ \hline
Schneider Electric         & CVE-2022-45789 \\ \hline
\end{tabular}
\end{table}


\vul{Remote Code Execution (RCE)} allows an attacker to execute arbitrary code on a device.
We identified nine RCE vulnerabilities listed in Table~\ref{rce}.
%
%
%
%
%
We explain the two main pathways to gaining RCE \ie logic downloads and firmware updates.

\begin{table}[h]
\caption{RCE}
\label{rce}
\centering
\begin{tabular}{|p{0.22\linewidth}|p{0.68\linewidth}|}
\hline
\textbf{Vendor} & \textbf{CVEs}                  \\ \hline\hline
Bently Nevada   & CVE-2022-29953                 \\ \hline
Emerson         & CVE-2022-30264, CVE-2022-30265 \\ \hline
Honeywell       & CVE-2022-30315                 \\ \hline
JTEKT           & CVE-2022-29958                 \\ \hline
Motorola        & CVE-2022-30269                 \\ \hline
Omron           & CVE-2022-31206                 \\ \hline
Phoenix Contact & CVE-2022-31800, CVE-2022-31801 \\ \hline
Schneider Electric  & CVE-2022-45788 \\ \hline
\end{tabular}
\end{table}

\paragraph{RCE via Logic Downloads}
%
We examined the logic generation and runtime mechanisms used in each
device, spanning over a total of 10 different runtime systems. 
We found that the vast majority compile their logic to native machine code for direct execution on the CPU module’s microprocessor. 
The dominance of native machine code execution is worrying given that it is by far the easiest target for achieving RCE.
This is compounded by the fact that none of the examined systems signed their logic, and except for one product family, no sandboxing was used for those systems that executed native machine code. 
In addition, many devices that we examined, use hardware and OS combinations which do not allow memory and privilege separation, resulting in attacker code execution with the highest possible privileges.

One of the most prominent defenses against unauthorized logic downloads is relying on physical mode switches governing controller operating modes.
These switches make a distinction between RUN and PROGRAM modes so that operators can follow explicit policy to only ever switch to PROGRAM mode during work-order initiated changes.
We found that only 17\% of devices include distinct RUN and PROGRAM modes.
It is noteworthy that even in the case of switches which do distinguish between RUN/PROGRAM modes, the absence of logic signing means that an attacker can still strike during legitimate maintenance windows.
Therefore, in a non-trivial number of cases, attackers who merely figure out how a logic download is performed against a system can achieve unconstrained code execution unhindered by any physical security measures.


\paragraph{RCE via Firmware Updates}
%
We investigated the firmware updating mechanisms of each device,\footnote{Five product families were not applicable for this analysis.} and
we found that the vast majority \ie 62\% transmit firmware updates to OT devices via Ethernet.
The other channels were SD cards (12\%), serial connections (10\%), USB (8\%), and 8\% of products adopted a combination thereof.
Only 51\% of the examined devices had some sort of authentication for firmware updates, even if it was in the form of hardcoded credentials in some cases.
The majority \ie 78\% lacked any sort of cryptographic firmware signing.
The findings are worrying because they suggest that many firmware update mechanisms are insecure, either by design or implementation and they are also often exposed to the network.
%
%
Even though serial, USB, and SD card transmission channels are less risky than Ethernet-based ones, it is important to remember that without proper firmware signing, these devices are still at risk of being compromised by engineering workstations and attackers piggybacking on legitimate firmware updates.
%
%
%




\subsection{Impact}

Several of the vulnerabilities uncovered are considered critical.
We estimated where the vulnerable systems are used to shed more light on the impact of these vulnerabilities.\footnote{We used sources that can contain false positives and none of them is exhaustive.}
 
%

%
%
We looked at product documentation, datasheets and marketing information that mention where they are used and for what purposes.
We found that water and wastewater, oil and gas, and power generation are the three common sectors that use these OT products.
%
%
We also queried Forescout Device Cloud for the vulnerable devices and found close to 30 thousand results.\footnote{Forescout Device Cloud is a repository of information of 18+ million devices monitored by Forescout appliances present in customer networks.}
Manufacturing was the top sector with 26\% of devices, followed by healthcare (16\%), retail (14\%), and government (12\%).
There were only a small presence in Oil and Gas sectors, but that is likely because many of those types of customers do not share device information with Forescout’s Device Cloud.
%
%
We also used Shodan search engine to check whether any of the vulnerable devices are connected to Internet.\footnote{https://www.shodan.io}
Estimating the number of affected devices based on public data is difficult because these devices are not supposed to be accessible via the internet.
However, we were still able to see a few thousands that are exposed online.
We found that the top most exposed products are 
Honeywell Saia Burgess, Omron, and Phoenix Contact with 2924, 1305, and 705 records respectively.
Geolocation information provided by Shodan indicates that a significant number of the exposed devices are located in Europe.
Italy is the first with a total of 1255 exposed devices, followed by Germany (440), Spain (393), and France (376).
Table~\ref{shodan} lists the number of top three exposed devices in the above countries. 
We also found 263 exposed devices in Switzerland and 178 devices in the US.

\begin{table}[]
\centering
\caption{The top most exposed devices}
\label{shodan}
\begin{tabular}{l|l|l|l|l|}
\cline{2-5}
\multicolumn{1}{r|}{}                        & Italy & Germany & Spain & France \\ \hline
\multicolumn{1}{|l|}{Honeywell Saia Burgess} & 954   & 326     &       & 192    \\ \hline
\multicolumn{1}{|l|}{Omron}                  &       &         & 321   & 110    \\ \hline
\multicolumn{1}{|l|}{Phoenix Contact}    & 285   & 104     & 55    &        \\ \hline
\end{tabular}
\end{table}

\section{Misleading Certification}

We noted that
74\% of the products had achieved one or more of the following certifications at at least SL1:\footnote{There were also products certified by \eg ISASecure Secure Development Lifecycle Assurance (SDLA) and GE Achilles Practices Certification (APC). We did not consider them due to their implied nature in other certifications.}

\begin{itemize}  

\item ISASecure Component Security Assurance (CSA): Subsumes Embedded Device Security Assurance (EDSA), based on IEC 62443-4-1 and IEC 62443-4-2. 
\item ISASecure System Security Assurance (SSA): Based on IEC 62443-4-1 and IEC 62443-3-3.
\item GE Achilles Communications Certification (ACC): Similar to, but not based on, IEC 62443-4-2.
\item ANSSI Certification de Sécurité de Premier Niveau (CSPN): Based on Common Criteria.

\end{itemize}  

Unexpectedly, we also found several products that claimed to have a security posture based on ``IEC 62443" and in some cases up to ``SL3/SL4", but they were not certified as such.
In other words,
products that were IEC 62443 certified, suffered from issues which should have been caught by these standards.
This suggests that apart from what the standards may not cover, even the things they do cover are not always properly covered in practice.

Considering that the vulnerabilities discussed in this work are either the result of insecure-by-design or (often trivial) security design failures, there seem to be serious issues with security standards and certifications for OT.
%

\paragraph{\textbf{Opaque Security Definitions}.}
Many security standards use opaque definitions. 
For instance, the IEC 62443 Security Assurance Levels correspond to attacker levels of sophistication. 
However, this sophistication is defined in very generic and opaque terms such as ``moderate resources", ``sophisticated means" and ``IACS specific skills". 
These terms, when left vague and unquantified, lend themselves to idiosyncratic interpretations more reflective of the auditor's perceptions and expectations than of a product's security posture. 
For example, consider a decent Capture the Flag (CTF)-playing teenager with a few weeks of spare time, a free copy of Ghidra, dotPeek, Wireshark, and a Wikipedia-level understanding of Modbus.
It is unclear whether she is a level 3 or 4 attacker.
Indeed, such an adversary would be more than capable of reverse-engineering and exploiting most proprietary Modbus extensions.
To complicate matters further, the IEC 62443 security requirements are incremental.
That is, the requirements for level 1 are a subset of level 2 and so on. 
Therefore, a product evaluated for mere protection against ``unintentional misuse" (\ie level 1) will have an authentication requirement that satisfies protection against ``state-sponsored actors" (\ie level 4). 
What is worrying is the fact that once a product is certified at a certain level and aims to achieve the next, it is unlikely that auditors re-evaluate already met requirements with more scrutiny rather than just focusing on the missing requirements.

\paragraph{\textbf{Limited Target of Evaluation}}
Security certifications typically have an implicit or explicit Target of Evaluation (ToE) such as a specific system or set of its components.
Each evaluation examines a given set of security requirements according to a given security profile or level.
It is not uncommon for ToEs to be incredibly limited and not cover significant attack surface such as proprietary engineering functionality or third-party networking libraries.
We observed that some vendors have opted for certification only at the lowest level and some others claim to have developed products ``according to" a standard but have not gone through the effort of actual certification.
In addition, many security certification processes limit the evaluation of security requirements to functional testing.
That is, features are verified to be present but no inspection for robustness is made. 
%
%
This type of testing typically excludes any sort of investigation of proprietary protocols, and as such, a functional security assessment might conclude authentication is present on an engineering interface while in reality the protocol is unauthenticated and all authentication is done client-side. 
Similarly, fuzzing- and conformance-based communication robustness tests are able to assess open protocols for which the specifications are known by the auditors. 
In other words, even if there are standards that are not limited to functional testing and include security assessment, a lack of access to source code or detailed specifications of proprietary protocols require that auditors spend a significant time to reverse-engineer functionalities, which increases chances that these proprietary aspects are left out of the ToE.
The wide range of OT products with broken authentication and access control schemes (e.g., plaintext credential transmission, hardcoded keys, broken or weak cryptography)  that we reported in this study represent the limits of functional testing.
%

\section{Unsound Risk Management}

Vendors often recommend asset owners to focus on perimeter hardening and monitoring to protect OT devices against security risks.
This requires asset owners to know in what ways a device or protocol is at risk.
However, vendors are often disinclined to share such information and the onus has typically been on asset owners or third-party security experts to bridge this gap.
We discuss two hindrances that complicate sound risk management in this domain.



\paragraph{\textbf{Obscure Vulnerability Information}.}
We reported the vulnerabilities to vendors.
We observed several cases where vendors and CISA did not publish their advisories until we publicised the vulnerabilities. 
In some cases, the information was delayed for months over minor divergences about mitigation steps or CVSS scores, for instance.
Therefore,
relying on a centralized database of vulnerabilities that is mainly maintained by vendors (such as CISA or NIST's NVD for CVEs) does not guarantee timely information for asset owners.
We realized that some vendors consider insecure options that exist for legacy compatibility not to be a vulnerability.
For instance,
firmware images are not signed in Yokogawa STARDOM and only rely on insecure checksums for regular integrity checks. However, the vendor refused to assign a CVE ID for FSCT-2022-0039.
Likewise, Emerson considered CVE-2022-29954, CVE-2022-29955, CVE-2022-29956, CVE-2022-29961, CVE-2022-30262 not to be vulnerabilities and therefore did not issue advisories for those. 
Therefore, relying solely on vendor information is not only untimely but also incomplete, which leaves asset owners exposed to risks.
There was one vendor that chose to delay the disclosure for months because they did not have a patch ready for their issues. 
Hence, the untimely and incomplete nature of vulnerability information in OT/ICS stems not just from miscommunication or disagreements about advisories but also from vendors that prefer to withhold information until a ``perfect'' solution exists. 
Their usual reasoning is that announcing a vulnerability without a patch increases risk, but the fact is that the vulnerability remains in the product whether or not it is announced and the risk continues to exist in asset owners networks.
Without explicit CVEs, asset owners could be forgiven for prioritizing security efforts geared to addressing minor issues with CVEs over addressing more impactful issues for which there is no CVE or official guidance.


\newpage
\paragraph{\textbf{Supply-chain Issues}.}
We also encountered vulnerabilities on an important supply chain component of OT devices named ProConOS runtime system.\footnote{A runtime system is the component responsible for running the logic program of the PLC, which cyclically scans the device's input and produces the desired output.
ProConOS provides the backbone for programmable controllers of many different vendors allowing for the execution of IEC 61131 and C\# programs as well as various controller management functions.
}
%
%
Due to the lack of Software Bill of Materials and the complexity of product supply chains, it is often not immediately clear what runtime a particular PLC uses.
Runtimes are subject to OEM integration decisions.
That is, a PLC manufacturer may choose to use a runtime but not the protocols (preferring to use their own), or they may choose to use the protocol on a non-default port, or a manufacturer may even choose to rebrand or modify the runtime altogether.
Consequently, it is not immediately visible what other products are affected by an issue discovered in a supply chain component.
Examples are the two CVEs (CVE-2014-9195 and CVE-2019-9201) that were assigned to ProConOS protocols in the past.
These CVEs were associated only with Phoenix Contact and its own PLCs but not propagated to other vendors incorporating this software package. 
This means that various other vendors and controllers using ProConOS/eCLR and the asset owners using them are likely unaware of these security issues. 
This has already led to a duplicate finding in the past with CVE-2016-4860, which seems specific to Yokogawa STARDOM controllers and does not mention that it is the same issue as CVE-2014-9195.
In a similar vein,
we encountered ProConOS/eCLR runtime and its associated protocols (the SOCOMM, ADE, or DDI) several times, but we could not find any associated CVEs for these products or public discussion that they were affected.
Thanks to open-source intelligence and our internal research, we have compiled a more extensive overview of vendors and products that are at risk, which are present in Table~\ref{runtime-risks}.
We coordinated with Phoenix Contact, CERT VDE, and CISA to contact these downstream vendors individually and hopefully include them in the original CVEs.
In addition, while prior work has pointed out the possibility for remote code execution on ProConOS systems, no CVE was assigned for this. 
We have confirmed this possibility against many other devices, and we requested CVE-2022-31800 to cover this issue for all affected vendors.
It is important to note that not all products affected by CVE-2022-31800 are also affected by CVE-2014-9195 or CVE-2019-9201. 
For example, the Emerson ControlWave family of PLCs/RTUs uses the ProConOS runtime but performs engineering operations using its own proprietary BSAP/IP protocol.
Therefore, proactive propagation of supply chain vulnerabilities is necessary to prevent
CVE duplication and improve root-cause analysis.

\begin{table}
\caption{Products at risk for using the ProConOS/eCLR Runtime and/or the SOCOMM, ADE or DDI protocols}
\label{runtime-risks}
\begin{tabular}{|l|p{0.65\linewidth}|}
 \hline
 Vendor & Products \\ [0.5ex] 
\hline
\hline
Phoenix Contact & AXC 1xxx, AXC 3xxx, RFC 4xx, ILC 1xx ETH, ILC 3xx, FC 200, FC 350  \\
\hline
Emerson  & ControlWave ``Next Generation" such as CWM, CWP, CWX, CMR, CME, GFC, XFC, EFM, PAC and LP \\
\hline
ABB  & RTU 5xx (RTU520/RTU540/RTU560)  \\
\hline
Advantech  & ADAM-3600, ADAM-5xxx, APAX-5xxx, APAX-6xxx, AMAX-2050, UNO-2171  \\
\hline
KUKA  & KUKA.PLC  \\
\hline
ICP DAS  & KinCon-8xxx  \\
\hline
Yaskawa  & MPiec  \\
\hline
Schleicher  & XCx (300/400/500/700/800/1100/1200)  \\
\hline
Hilscher  & netPLC  \\
\hline
Luetze  & DIOLINE PLC  \\
\hline
Delta  & DMXC  \\
\hline
ISH  & SIS800, SIC400, uPLC iSOC300P  \\
\hline
Yokogawa  & STARDOM FCJ, FCN-RTU, FCN  \\
\hline
\end{tabular}
\end{table}


%
%

%
%
\section{Related Work}
\label{sec:relatedwork}



\paragraph{Fuzzing Specific Protocols}
%
%
Tacliad et al. developed a Scapy-based fuzzer to find potential vulnerabilities due to corrupted packets in EtherNet/IP implementations~\cite{Tacliad}.
They used remote fault detection methods to identify faults triggered by fuzz testing, and they discovered a new vulnerability in a common industrial controller used in Navy control systems which can cause Denial of Service (DoS).
%
%
%
%
Zhao et. al proposed SeqFuzzer, a framework that automatically learns the protocol frame structures from communication traffic and generates fake but plausible messages as test cases~\cite{Zhao}. They discovered a number of security vulnerabilities on widely used Ethernet for Control Automation Technology (EtherCAT) devices.
Tychalas et al. introduced IFFSET (InField Fuzzing through System Emulation Tool), a platform for security assessment of specific libraries inside ICS components and showed that it is effective in finding potential vulnerabilities commercial PLCs without affecting the control process~\cite{Tychalas}.
Niedermaier et al. introduced PropFuzz, a new stable and extensible framework to fuzz proprietary ICS protocols and track controller behavior~\cite{Niedermaier}. It can be used to analyze IDE and PLC communication, and fuz DuT, and the tool can track the output and detect any unusual activity. 
The authors applied PropFuzz to two PLCs and found three critical vulnerabilities that attackers may remotely exploit (Advisory ICSA-16-313-01).
Yu et al. proposed CovGAN, a generative adversarial network that generates test cases by learning from IIoT protocol specifications~\cite{Yu}. 
They built a fuzzing framework (CGFuzzer) and a protocol simulator based on CovGAN, and they showed that CGFuzzer outperforms GANFuzz, SeqFuzz, and Peach in terms of passing rate and code coverage.


\paragraph{Analysis of Single Device/Protocol}
Beresford et al. discussed how to attack a Siemens Simatic S7 PLC using reconnaissance, fingerprinting, replay attacks, authentication bypass techniques, and remote exploitation~\cite{Beresford}.
Biham et al. extended attacks that can remotely start or stop the PLC to the latest S7-1500 PLCs~\cite{Biham}. 
%
They downloaded a control logic of the attacker’s choice to a remote PLC, and were able to modify the running code and the source code, which are both downloaded to the PLC.
Hui et al. examined potential exploits of the Siemens S7-1211C controllers and the Totally Integrated Automation (TIA) engineering software~\cite{hui}. Using Windbg and Scapy, the anti-replay mechanism of the Siemens proprietary communication protocol, they found that S7CommPlus, and the Profinet Discovery and Basic Configuration Protocol are vulnerable to attacks such as session stealing, phantom PLC, cross-connecting controllers, and denial of S7 connections.
Finally, Abbasi et al. obtained unconstrained native kernel code execution through a physical access backdoor in the PLC, allowing for detailed reverse engineering and debugging~\cite{abbasi-2019}.
%


\newpage

\paragraph{Small-scale Analysis of Devices/Protocols}
Martin-Liras et. al developed a testbed for security assessment of vendor-specific protocols, but they only looked at three vendors and did not assign CVE IDs to their findings~\cite{Martin}. Wightman et. al focused on insecurity by design of PLC equipment that was a seminal moment in OT/ICS security research, but they only investigated five models and their research was done before a decade of real-world malware and OT incidents had happened~\cite{Wightman}. 
Figueroa-Lorenzo et al. listed 33 common protocols, standards and buses used in Industrial IoT and collected security issues that were published for each one~\cite{Figueroa-Lorenzo}.
%
Gonzalez et al. analyzed security advisories vulnerability reports and found that HMIs, SCADA, and PLCs have been most vulnerable ICS components, and that these products were designed with the expectation that they would be utilized in a secured network.~\cite{Gonzalez}.




%
%

\section{Conclusion}
\label{sec:conclusion}

We report 53 distinct CVEs related to 45 OT products from ten major vendors and uncover that every product suffers from at least one trivial vulnerability.
We conclude that despite a decade of efforts in improving OT security, the OT install base is still suffering from insecure-by-design issues even for products that are security certified.
Unlike other studies that have made similar points, we analyzed products actively deployed at large industrial and critical infrastructure customers. 
We also covered a wider range of vendors, systems, and sectors affected.

%


\bibliographystyle{ACM-Reference-Format}
\bibliography{reference}

\end{document}